\documentclass[english]{article} 
\usepackage{color}
\makeatletter
\usepackage[latin1]{inputenc}
\usepackage{graphicx}
\usepackage{hyperref}

\usepackage{geometry}
\usepackage{color}

\usepackage{amstext}
\usepackage{amssymb}
\usepackage{color}

\newcommand{\vx}{\vec x}
\newcommand{\vxp}{{\vec x}\,'}
\newcommand{\vk}{{\vec k}}

\makeatother

\begin{document}

\title{About non-relativistic quantum mechanics and electromagnetism.}

\author{Ladislaus Alexander B\'anyai*  and Mircea Bundaru**}

\maketitle

{\sl Institut f\"ur Theoretische Physik, Goethe-Universit\"at, Frankfurt am Main, Max-von-Laue-Strasse 1, D-60438, Germany}*

{\bf banyai@itp.uni-frankfurt.de}, {\color{blue}corresponding author} 

{\sl National Institute of Physics and Nuclear Engineering - Horia Hulubei, PO-Box MG-6, Bucharest-Magurele, Romania}**

{\bf bundaru@theory.nipne.ro}

\abstract{We describe here the coherent formulation of electromagnetism in the non-relativistic quantum-mechanical many-body theory of interacting charged particles. We use the mathematical frame of the field theory and its quantization in the spirit of the QED. This is necessary because a manifold of misinterpretations emerged especially regarding the magnetic field and gauge invariance. The situation was determined by the historical development of quantum mechanics, starting from the Schr\"odinger equation of a single particle in the presence of given electromagnetic fields, followed  by  the many-body theories of many charged identical particles having just Coulomb interactions. Our approach to the non-relativistic QED  emphasizes the role of the gauge-invariance and of the external fields. We develop further the $1/c^2$ approximation of this theory allowing a closed description of the interacting charged particles without photons. The resulting Hamiltonian coincides with the quantized version of the Darwin Hamiltonian containing besides the Coulomb  also a current-current diamagnetic interaction. We show on some examples the importance of this extension of the many-body theory.}


\section{Introduction}

The quantum mechanics of a single charged particle in the presence of applied electric  and magnetic fields is based on the classical  mechanics of a point-like charge derived from a Lagrange function, followed by a quantization according to  the "Poisson-bracket into commutator" recipe. This scheme  however failed  for the  electromagnetic interaction between charged particles. The classical electromagnetic theory of point-like charges has neither a Lagrangian nor Hamiltonian formulation. One eliminates by hand from the Lorentz force the action of the fields created by the particle itself. Nevertheless, one took over this recipe in order to include at least the Coulomb interaction between the particles.  The constructed Hamiltonian lead to big successes as well in the theory of atomic structure, as in the solid-state theory. Of course, one had to include also the symmetry or anti-symmetry of the wave functions due to the bosonic, or fermionic nature of identical particles, as well as the spin, whose origin stems actually from the relativistic theory. This state of the solid-state theory however was unable to include the magnetic field and forces created by the charged particles themselves. The importance of this problem was waved out by an argument relying on the small velocities of the electrons and ions in matter as to be compared to the light velocity $c$. This argumentation is however false, since  it is well-known that a slow flow of a macroscopic number of electrons may create enormous magnetic fields.

 A first attempt to construct a Lagrangian and Hamiltonian including  velocity-dependent forces up to $1/c^2$ in the classical theory of interacting charged point-like particles was made by Darwin Ref.\cite{Darwin} 100 years ago. His idea was to neglect the retardation  from the Lenart-Wichert potentials in order to eliminate them in favor of the particle velocities and ignoring the terms containing the velocity of the same particle. Later Landau and Lifshitz  Ref. \cite{Landau} as well as   Jackson \cite{Jackson} realized, that one may not pursue this way without fixing the gauge. Both used implicitly or explicitly the Coulomb gauge in their derivation. This approach based mostly on physical intuition still does not offer a sound basis for a theory of interacting charged particles, since the so obtained Hamiltonian again cannot be integrated in  a correct Lagrangian formulation due to the nonlinear velocity dependent forces it implies, as well as to the artificial exclusion of the self-interaction.  It is therefore not surprising that the Darwin model had no impact on the quantum-mechanical many-body theories of condensed matter developed in the second half of the 20th century. 

 This is a very strange situation, since already in the mid 20th century Quantum Electrodynamics (QED) was successfully developed  and its is considered as the basic theory for  the relativistic, quantum mechanical description of electrons and positrons interacting   with photons. 
 Therefore, at least a non-relativistic quantum-mechanical description of electromagnetic interacting electrons is desirable  as an approximation to the QED. Such a reduction however seems illusory and again  leads into a deadlock, since without positive charges no matter is stable. The condensed matter we have to  deal is  made up of electrons and nuclei (not positrons!).  The nuclei are not elementary particles in the present understanding of the quantum field theory, but composite objects made out of protons an neutrons bound together by non-electromagnetic forces. Even worse, the protons and neutrons themselves are not the ultimate basic pieces. A  certain skepticism is justified  regarding  a derivation of  condensed matter theory from the standard model of interacting elementary particles.

After renouncing to the ill-defined classical theory of point-like charges one has to build a consistent electromagnetic theory of arbitrary charged quantum mechanical particles by following the mathematical pattern of the QED. Such a theory is compulsory non-relativistic, since it does not include antiparticles. Quantization of relativistic wave equations however implies antiparticles!

The non-relativistic QED (including the relativistic concept of the spin and its magnetic moment) was not only formulated, but also widely used in quantum optics, where one treats photons interacting with atoms, molecules or solids.  (See for example a recent presentation in Ref. \cite{Koch}.) 

In the same time, most theories of condensed matter still  use many-body formulations including only Coulomb forces between the charged particles, while one avoids the original configuration-space formulation that excludes self-interaction. Nevertheless,  by  the second quantization  on introduces  self-interaction implicitly again.  

This rather unsatisfactory situation is due partly to the wide  split between the solid state theorists and  elementary particle theorists. 
The purpose of this paper  (partly discussed in the book \cite{Lectures2}) is to provide a clear Lagrangian presentation  of the non-relativistic QED of charged particles and within this frame the  derivation    of a consistent reduced Hamiltonian valid up to order $1/c^2$ containing only the electromagnetically interacting charged particles and no photons.   
This derivation \cite{Banyai1} is based essentially on the old paper of Holstein, Norton and Pincus \cite{Holstein} about the transverse photon exchange being at the origin of diamagnetism.
It occurs that this Hamiltonian is nothing else as the quantum mechanical version of the  Darwin Hamiltonian \cite{Darwin} in Jackson's formulation \cite{Jackson}.

We start with  the classical  field theory of electromagnetic fields interacting with a  Schr\"odinger electron obeying the Maxwell equations. This approach is analogous to the starting point of the relativistic QED, which is based on the Dirac equation. The next step is the construction of the Lagrangian, which is not uniquely defined. 

An essential role in any field theory play the external sources. Without these macroscopic, classical entities, no measurement  may be considered.  Unfortunately, a certain confusion is still  spread in the literature about the internal and external electromagnetic fields.( See however the 50 years  old book of Zubarev \cite{Zubarev}, where he already insisted on the importance of the general distinction between external and total fields in the frame of the linear response.)   We insist here on this very important aspect since it is essential for the interpretation of experiments. 

There are two different ways to introduce the external fields or sources. 
The first considers the "internal" e.m. fields produced by the electron that interacts also with "external" fields (produced by external sources) by the minimal e.m. coupling and it gives rise to an explicitly gauge invariant  Lagrange density. This version is used to build up a Hamiltonian.  Since the Lagrangian is "singular" one needs to fix the gauge, namely by choosing the Coulomb one. This choice of gauge eliminates the spurious degrees of freedom, allowing to avoid the problems raised by Dirac's theory of canonical formalism \cite{Dirac1,Dirac2} in the presence of constraints. The quantization of this Hamiltonian leads to the usual non-relativistic QED. The gauge freedom here is restricted to the transformations of the external fields. This aspect is ignored in all the discussions of magnetic effects in solid state.

The second way is to introduce explicitly a coupling of the e.m.  fields directly  to the external sources. The resulting electromagnetic fields are the "total" ones. The external electromagnetic potentials are eliminated in  favor of their sources. The Lagrange density is not any more gauge invariant, but the action is still gauge invariant.  This way is used in the functional integral formulation of the QED,  which is defined just by the Lagrange density. No Hamiltonian is needed at all. The two approaches are equivalent, at least for simple connected systems.

Thereafter we discuss the quantization of the first variant  i.e. the non-relativistic QED in the Hamiltonian formulation and include also  the ingredient of the spin-magnetic moment taken over from the relativistic QED. 

We describe also the  $1/c^2$ approximation of the non-relativistic QED   in the absence of photons suitable for condensed matter theory. Within this frame already the magnetic interaction between the charged particles is taken into account.

While formally we considered here just "electrons" the theory may be extended trivially to a system of negatively and positively  charged particles (electrons with ions or electrons in conduction band with holes in the valence band).              

It is worth to mention here, that unlike in the relativistic QED, the divergences in the adiabatic perturbation theory of the S matrix were not yet treated systematically. The many-body theory of solid-state is used always in its cut-off form and not by a renormalization of the bare parameters.  Helpful here are the "natural" cut-off parameters like the Debye wave-length or the band width in a crystal.

Further, we describe the important role of this improved many-body theory for electromagnetic linear response, superconductivity and thermal noise. 
The transverse dielectric function (or conductivity) is meaningless in a pure Coulomb theory. Only the magnetic field created by the electrons may explain  perfect diamagnetism shown in the Meissner effect. The thermal noise of electromagnetic origin must distinguish between the longitudinal /transverse fields or photon number measurements.
 All these fields need to consider magnetic forces between the charged particles, at least in the $1/c^2$ approximation.

\section{Classical field theory.}
In the classical field-theory one defines the action ${\cal A}$ 
\begin{equation}
{\cal A}=\int d {\vec x} \int dt {\cal L}({\vec x},t) \label{action}
\end{equation}
by a Lagrange density ${\cal L}({\vec x},t)$ depending on some fields $\phi_i({\vec x},t)$  together with their first time and space derivatives. The variational principle $\delta A=0$ gives rise to the generalized Euler-Lagrange equations 

\begin{equation}
\frac{\partial}{\partial t}\frac{\delta {\cal L}}{\delta {\dot \phi_i({\vec x}, t)}}+
\frac{\partial}{\partial x_\mu} \frac{\delta {\cal L}}{\delta 
\frac{\partial \phi_i({\vec x}, t)}{\partial x_\mu} }
-\frac{\delta {\cal L}}{\delta \phi_i({\vec x},t)}=0 \enspace .
\label{Euler-Lagrange}
\end{equation}
Here the symbol $\partial$ means ordinary derivative, while the symbol $\delta$ means functional derivative.  Two Lagrangian densities that differ by the time derivative  or by the divergence of a function give rise to the same action and therefore are considered to be equivalent, taking into account the vanishing of the fields at infinity.

The generalized canonical conjugate momenta for the fields $\phi_i ({\vec x},t)$ are defined by 
\begin{equation}
\Pi_{\phi_i} =\frac{\delta{\cal L}}{\delta {\dot {\phi}_i}}
\end{equation}
and the Hamiltonian density  is
\begin{equation}
{\cal H}(\phi,\Pi_\phi)=-{\cal L}+\Pi_{\phi_i} {\dot \phi_i} \enspace ,
\end{equation}
provided no relations (constraints) appear between the canonical conjugate momenta. Lagrangians with constraints however have to be handled with Dirac's canonical formalism \cite{Dirac1, Dirac2}, that implies also a redefinition of the Poisson bracket. 

\section{Classical Maxwell fields coupled to a quantum mechanical electron and external sources.}

The  classical Maxwell equations  are two with sources $\rho$ and
 $\vec{j}$
\begin{eqnarray}
\nabla\times\vec{B} & = & \frac{4\pi}{c}\vec{j}+\frac{1}{c}\frac{\partial}{\partial t}\vec{E} \label{Max1a}
\\
\nabla\vec{E} & = & 4\pi\rho \label{Max2a}
\end{eqnarray}
and two  without sources 
\begin{eqnarray}
\nabla\vec{B} & = & 0\label{Max3a}
\\
\nabla\times\vec{E} & = & -\frac{1}{c}\frac{\partial}{\partial t}\vec{B} \enspace 
\label{Max4a}
\end{eqnarray}
(we use Gaussian units like Ref. \cite{Landau} ).
The equations without sources are automatically satisfied by the introduction of the  electromagnetic potentials
 
\begin{eqnarray}
\vec{B} & = & \nabla\times\vec{A} \label{e.m.pot1}
\\
\vec{E} & = & -\nabla V-\frac{1}{c}\frac{\partial}{\partial t}\vec{A} \enspace .
\label{e.m.pot2}
\end{eqnarray}
Therefore we may concentrate on Eqs. \ref{Max1a} and \ref{Max2a} depending on the sources. 

We are looking for the "internal" electromagnetic fields produced by a single quantum mechanical electron.
Thus the sources $\rho$ and ${\vec j}$ are given by the  
quantum mechanical charge and  current of an electron described by the wave function   $\psi({\vec x},t) $ (for simplicity without spin) coupled to the ("internal") electromagnetic fields $\vec{A}$, $V$ as well as to some "external" classical fields $\vec{A}_{ext}$,
$V_{ext}$. The later are supposed to satisfy their own Maxwell equations with the external macroscopic sources  
$\rho_{ext}$ and $\vec {j}_{ext}$. 

Thus
\begin{eqnarray}
\rho({\vec x},t)&=&e\psi({\vec x}, t)^* \psi({\vec x},t) \label{source1}
\\
\vec{j}({\vec x},t)&=&\frac{e}{2m}\psi({\vec x},t)^*\left(-\imath \hbar\nabla
+\frac{e}{ c}({\vec A}({\vec x},t)+{\vec A}_{ext}({\vec x},t)) \right)\psi({\vec x},t) +c.c  \label{source2}
\end{eqnarray}
while the  wave function $\psi({\vec x},t)$ satisfies the Schr\"odinger equation
\begin{equation}
\imath\hbar\frac{\partial}{\partial t}\psi({\vec x},t)=
\left(\frac{1}{2m}\left(-\imath\hbar\nabla+\frac{e}{c}(\vec{A}(x,t)+\vec{A}_{ext})\right)^{2}
+e\left(V({\vec x},t)+ V_{ext}(\vec{x},t)\right)\right)\psi({\vec x},t) \enspace . \label{Schro}
\end{equation}

Using only this equation one gets the continuity equation

\begin{equation}
\nabla\vec{j}+\frac{\partial}{\partial t}\rho=0 \label{continuity}
\end{equation}
 required for the consistency of Eqs. \ref{Max1a} and \ref{Max2a}.

The Schr\"odinger equation that couples to the e.m. fields, as well as the current density  were introduced here according to the minimal recipe
\begin{eqnarray}
\frac{\hbar}{\imath}\nabla\psi & \to & \left(\frac{\hbar}{\imath}\nabla +\frac{e}{c}({\vec A}+{\vec A}_{ext})\right) \psi \label{minim1}
\\
\frac{\hbar}{\imath}\frac{\partial}{\partial t} \psi &\to &  
\left( \frac{\hbar}{\imath}\frac{\partial}{\partial t} + 
e(V+eV_{ext})\right))\psi \enspace .
\label{minim2}
\end{eqnarray} 
This is a peculiar case of the covariant (here abelian) Yang-Mills derivative.

An alternative formulation is to define the "total" electromagnetic potentials and fields  as those produced by the electron and the external sources
\begin{equation}
\vec{A}'\equiv \vec{A}+\vec{A}_{ext}; \qquad
V'\equiv V+V_{ext}
\enspace , \label{total} 
\end{equation}
which are usually denoted with the same symbol as the internal ones that may lead to confusions. Their sources are given  by the sum of the  electron charge and current densities and the external charge and current densities. Then no external fields, but just their sources appear in the theory. However it is very important to discern the two descriptions, although formally they are equivalent.

The electromagnetic potentials and the wave function are not uniquely defined, they allow a simultaneous gauge  transformation 
\begin{eqnarray}
V({\vec x},t)&\to& V({\vec x},t)+\frac{1}{c}{\dot \chi({\vec x},t)} \label{gauge}
\\
\vec{A}({\vec x},t)&\to& \vec{A}({\vec x},t) -\nabla \chi({\vec x},t)\nonumber
\\
\psi({\vec x},t) &\to& \psi({\vec x},t)e^{-\frac{\imath e}{\hbar c}\chi({\vec x},t)}\nonumber
\end{eqnarray}
with an arbitrary differentiable  function $\chi({\vec x},t)$ that do not change neither the Maxwell fields ${\vec B}$, ${\vec E}$, their sources $\rho$, ${\vec j}$,  nor the whole system of equations from Eq.\ref{Max1a} to  Eq.\ref{continuity}. 
A similar gauge invariance is true with respect to the independent gauge transformations of the external potentials. 

This arbitrariness is not unexpected, since we actually introduced 4 degrees of freedom instead of the 3 independent degrees of freedom (the transverse magnetic field and the longitudinal electric field).

\section{Classical Lagrange density}

The first problem is then to find a Lagrangian giving rise to these equation in terms of the fields $V({\vec x},t)$, ${\vec A}({\vec x},t)$ and $\psi({\vec x},t)$ as
 dynamical variables (generalized coordinates).

It is easy to see, that  the uncoupled Maxwell 
\begin{eqnarray}
\nabla\times\vec{B} & = & \frac{1}{c}\frac{\partial}{\partial t}\vec{E} \label{Max1a0}
\\
\nabla\vec{E} & = & 0 \label{Max2a0}
\end{eqnarray}
and Schr\"odinger equations
\begin{equation}
\imath \hbar \frac{\partial}{\partial t} \psi({\vec x},t)=-\frac{\hbar^2}{2m}\nabla^2\psi({\vec x}, t) \label{Schr0}
\end{equation}
 follow from the free Lagrangian density

\begin{eqnarray}
{\cal L}_{0} ({\vec x},t)&=&\frac{1}{8\pi}\left(\nabla V({\vec x},t)
+ \frac{1}{c} {\dot{\vec A}}({\vec x} ,t)\right)^2 -\frac{1}{8\pi}\left(\nabla \times {\vec A}({\vec x}, t)\right)^2 \label{Lfree}
\\
&-&\frac{\hbar^2}{2m}\nabla \psi^*({\vec x},t)\nabla \psi({\vec x},t) -
\frac{\imath\hbar}{2}\left({\dot \psi}({\vec x},t)^*\psi({\vec x},t)-\psi({\vec x},t)^*{\dot{\psi}}({\vec x},t)\right) \enspace .\nonumber 
\end{eqnarray}

The introduction of the e.m. fields by the minimal recipe Eqs. \ref{minim1} and \ref{minim2} into the free Lagrangian ${\cal L}_0$ leads to the Lagrange density
\begin{eqnarray}
{\cal L}({\vec x},t)&=&\frac{1}{8\pi}\left(\nabla V+ \frac{1}{c}\frac{\partial}{\partial t} {\vec A}\right)^2 -\frac{1}{8\pi}\left(\nabla \times {\vec A}\right)^2 \label{Lagran}
\\
&-&\frac{1}{2m}\left(-\frac{\hbar}{\imath}\nabla +\frac{e}{c}({\vec A}+{\vec A}_{ext})\right)\psi^*
\left(\frac{\hbar}{\imath}\nabla +\frac{e}{c}({\vec A}+{\vec A}_{ext})\right)\psi \nonumber
\\
&-&\frac{1}{2}\psi^*\left(\frac{\hbar}{\imath}\frac{\partial}{\partial t} +e(V +V_{ext}\right)\psi-
\frac{1}{2}\psi\left(-\frac{\hbar}{\imath}\frac{\partial}{\partial t} + e(V+V_{ext})\right)\psi^* 
\nonumber \enspace ,
\end{eqnarray}
that on his turn gives rise to the Maxwell Eqs. \ref{Max1a} and \ref{Max2a} as well as to the Schr\"odinger equation \ref{Schro}.
This Lagrange density  ${\cal L}$ is by construction invariant against  gauge transformations of the fields ${\vec A}$, $V$ and/or 
  ${\vec A}_{ext}$, $V_{ext}$.

On the other hand, if one works with the "total" potentials $\vec{A}'$, $V'$  of Eq. \ref{total} the Lagrange density looks as
\begin{eqnarray}
{\cal L}'({\vec x},t)&=&\frac{1}{8\pi}\left(\nabla V'+ \frac{1}{c}\frac{\partial}{\partial t} {\vec A}'\right)^2 -\frac{1}{8\pi}\left(\nabla \times {\vec A}'\right)^2 \label{Lagran1}
\\
&-&\frac{1}{2m}\left(-\frac{\hbar}{\imath}\nabla +\frac{e}{c}{\vec A}'\right)\psi^*
\left(\frac{\hbar}{\imath}\nabla +\frac{e}{c}{\vec A}'\right)\psi \nonumber
\\
&-&\frac{1}{2}\psi^*\left(\frac{\hbar}{\imath}\frac{\partial}{\partial t} +eV'\right)\psi-
\frac{1}{2}\psi\left(-\frac{\hbar}{\imath}\frac{\partial}{\partial t} + eV'\right)\psi^*  
\nonumber
\\
&+&  V'({\vec x},t)\rho_{ext}({\vec x},t)
+\frac{1}{c}{\vec A}'({\vec x},t){\vec j}_{ext}({\vec x},t) 
\nonumber\enspace .
\end{eqnarray}

This alternative Lagrange density is not explicitly  gauge invariant, but the corresponding action
 Eq.\ref{action} is still gauge invariant.  This may be shown after partial integration by  using the continuity equation of the external sources.   

This version is used in the functional (path) integral formulation of the QED in the whole homogeneous space to define the generating functional of the Greeen functions.  This formulation of the QED uses only the Lagrangian and needs no definition of any Hamiltonian.
 The two versions are however not completely equivalent.  
In what follows we shall develop a Hamiltonian formalism in the Fock space starting  from the Lagrangian density Eq. \ref{Lagran}  that is manifestly gauge invariant. This formulation is appropriate to include boundary conditions and multiple-connectivity.

\section{The classical Hamiltonian in the Coulomb gauge.}

In this Section  we build the classical Hamiltonian out of the Lagrangian Eq. \ref{Lagran}, needed for an operator formulation of the non-relativistic QED.
Unfortunately, this Lagrangian density is a so called singular one. The time derivative of the variable $V$ is not present in them and therefore the corresponding canonical momentum is vanishing i.e. we have a constraint in the canonical formalism. Lagrangians with constraints, as we already mentioned, have to be handled with Dirac's canonical formalism \cite{Dirac1, Dirac2}, that implies also a redefinition of the Poisson bracket.  
The simplest way out is however to use the choice of the gauge in such a way as to eliminate the spurious degrees of freedom from the Lagrangian before we could construct a Hamiltonian.

The Coulomb gauge defined by 
\begin{equation}
\nabla {\vec A}({\vec x},t)=0
\end{equation} 
leaves only the physical transverse degrees of freedom of the photons and simultaneously  eliminates the scalar potential in favor of the electron charge density 

\begin{equation}
V({\vec x},t)=\int d {\vec x}' \frac{\rho({\vec x}',t)}{|{\vec x}-{\vec x}'|} \enspace . \label{Coul}
\end{equation}

We shall construct the Hamiltonian the usual way, however taking into account of the above constraint and define the canonical conjugate momenta as 

\begin{eqnarray}
\Pi_{\psi}&\equiv & \frac{\delta  {\cal L}}{\delta {\dot \psi}} =\frac{\imath \hbar}{2} \psi^*
\\
\Pi_{\psi^*} &\equiv& \frac{\delta { \cal L}}{\delta {\dot \psi^*}} =-\frac{\imath \hbar}{2} \psi
\\
\Pi^{\mu}_{{A}_\mu}&\equiv & 
\frac{{\cal L}}{\delta {\dot A}^\mu }=
\frac{1}{4\pi c}\left(\frac{\partial}{x_\mu} V +\frac{1}{c} {\dot{ A}}^\mu\right); 
\quad (\mu=1^,2,3) 
\\
\Pi_V &\equiv& 0  \enspace .
\end{eqnarray}
and the Hamiltonian density  is
\begin{equation}
{\cal H} = -{\cal L}+ {\vec \Pi}_{{\vec A}} \dot{\vec  A} + \Pi_\psi {\dot\psi} +
\Pi_{\psi^*} {\dot \psi^*} \enspace .
\end{equation}

Using the notation $\vec{A}_\bot$ for the  vector potential in the Coulomb gauge we have explicitly
\begin{eqnarray}
{\cal H}
 &=&-\frac{1}{8\pi}\left(\nabla V+ \frac{1}{c} {\dot{\vec A}_\bot}\right)^2 +\frac{1}{8\pi}\left(\nabla \times {\vec A}_\bot \right)^2
+\frac{1}{4\pi c}{\dot {\vec A}_\bot} \left(\nabla V+ \frac{1}{c} {\dot{\vec A}\bot}\right)
\\
&+&\frac{1}{2m}\left(-\frac{\hbar}{\imath}\nabla \psi^* +\frac{e}{c}({\vec A}_\bot +{\vec A}_{ext})\psi^*\right)
\left(\frac{\hbar}{\imath}\nabla\psi +\frac{e}{c}({\vec A}_\bot+{\vec A}_{ext})\psi\right) 
+e(V+V_{ext})\psi^*\psi  \nonumber
\end{eqnarray}
or

\begin{eqnarray}
{\cal H}& =&
\frac{1}{8\pi}\left(\nabla V+ \frac{1}{c} {\dot{\vec A}_\bot}\right)^2 +\frac{1}{8\pi}\left(\nabla \times {\vec A}_\bot \right)^2
-\frac{1}{4\pi }\nabla V \left(\nabla V+ \frac{1}{c} {\dot{\vec A}_\bot}\right)
\\
&+&\frac{1}{2m}\left(-\frac{\hbar}{\imath}\nabla \psi^* +\frac{e}{c}({\vec A}_\bot+{\vec A}_{ext} )\psi^*\right)
\left(\frac{\hbar}{\imath}\nabla\psi +\frac{e}{c}({\vec A}_\bot+{\vec A}_{ext})\psi\right) 
+e(V+V_{ext})\psi^*\psi   \enspace . \nonumber
\end{eqnarray}
In   the  Hamiltonian
\begin{equation}
H\equiv \int d {\vec x} {\cal H}({\vec x})
\end{equation}
 one may use a partial integration in order to obtain

\begin{eqnarray}
H &=&\int d{\vec x}\left[\frac{1}{8\pi}\left(\nabla V+ \frac{1}{c} {\dot{\vec A}_\bot}\right)^2 +\frac{1}{8\pi}\left(\nabla \times {\vec A}_\bot \right)^2 
+\frac{1}{4\pi} V\nabla \left(\nabla V+ \frac{1}{c} {\dot{\vec A}}\right)
\right. 
\\
&+&\left.\frac{1}{2m}\left(-\frac{\hbar}{\imath}\nabla \psi^* +\frac{e}{c}({\vec A}_\bot +\vec{A}_{ext} )\psi^*\right)
\left(\frac{\hbar}{\imath}\nabla\psi +\frac{e}{c}({\vec A}_\bot +\vec{A}_{ext})\psi\right) + e(V +V_{ext})\psi^*\psi   \right]  \enspace . \nonumber
\end{eqnarray} 
 
Due to the transversality of the vector potential and expressing  the scalar  potential through the total charge density Eq. \ref{Coul} one gets further 
\begin{eqnarray}
H &=&\int d{\vec x}\left[\frac{1}{8\pi}\left(\nabla V+ \frac{1}{c} {\dot{\vec A}_\bot}\right)^2 +\frac{1}{8\pi}\left(\nabla \times {\vec A}_\bot \right)^2 \right.
\\
&+&\left.\frac{1}{2m}\left(-\frac{\hbar}{\imath}\nabla \psi^* +\frac{e}{c}({\vec A}_\bot +\vec{A}_{ext} )\psi^*\right)
\left(\frac{\hbar}{\imath}\nabla\psi +\frac{e}{c}({\vec A}_\bot +\vec{A}_{ext})\psi\right)+ eV_{ext}\psi^* \psi\right]  \enspace , \nonumber
\end{eqnarray}

or
\begin{eqnarray}
H\!&=&\!\!\int\!\! d{\vec x}\left[ \frac{1}{8\pi}\!\left({\vec E}^2 +{\vec B}^2\right)
+  \frac{1}{2m}\!\left(-\frac{\hbar}{\imath}\nabla \psi^*
 +\frac{e}{c}({\vec A}_\bot+{\vec A}_{ext})\psi^*\right)
\left(\frac{\hbar}{\imath}\nabla\psi +\frac{e}{c}({\vec A}_\bot+{\vec A}_{ext})\psi\right)\right.  \nonumber
\\
&+& \left.V_{ext}\psi^* \psi \right] \enspace . 
\end{eqnarray}     

Apparently the Coulomb interaction $eV\psi^*\psi $ disappeared, but actually it is contained properly in the
 energy  of the longitudinal electric field. Since under the integral
 \[ {\vec E}^2=(\nabla V)^2 + (\frac{1}{c}{\dot{\vec A}_\bot})^2 
 \]
 and after a partial integration
 \[ 
 (\nabla V)^2 \to -V\nabla^2V  
\]
 we get

\begin{eqnarray}
H &=& \int d{\vec x}\left[ \frac{1}{8\pi}\!\left({\vec E}_\bot^2 +{\vec B}^2\right) + e\psi^*\psi( \frac{1}{2} V +V_{ext})\right.  \label{Hamiltoniana-Coul}
\\
&+& \left.\frac{1}{2m}\left(-\frac{\hbar}{\imath}\nabla \psi^* +\frac{e}{c}({\vec A}_\bot +{\vec A}_{ext})\psi^*\right)
\left(\frac{\hbar}{\imath}\nabla\psi +\frac{e}{c}({\vec A}_\bot +{\vec A}_{ext})\psi\right) \right] 
\enspace . \nonumber
\end{eqnarray}
The first  term represents the energy of the transverse "photon" (radiation-) field. 

 Obviously, while the "internal" potentials are already fixed, one has still a  restricted gauge invariance of the Hamiltonian against the gauge transformations of the external potentials.

\section{Quantization.}
 Starting from our classical Hamiltonian in Coulomb gauge Eq.\ref{Hamiltoniana-Coul},
after the usual equal-time quantization of  the anti-commuting electron wave functions
\[
[\psi({\vec x},t),\psi^+({\vec x}',t)]_+=\delta({\vec x}-{\vec x}')
\]
 and the introduction of  creation and annihilation operators
  $b_{\vec{q},\lambda}^{+}$ and $b_{\vec{q},\lambda} $ of photons of polarization
   $\lambda$ and momentum ${\vec q}$ one  defines the quantized transverse e.m. vector potential 
 
\begin{equation}
\vec{A}_{\bot}(\vec{x})=\sum_{\lambda=1,2}\sqrt{\frac{\hbar c}{\Omega}}\sum_{\vec{q}}\frac{1}{\sqrt{|\vec{q}|}}\vec{e}_{\vec{q}}^{(\lambda)}
e^{-\imath\vec{q}\vec{x}}\left(b_{\vec{q},\lambda}+ b_{-\vec{q},\lambda}^{+}\right)
\label{Aquant}
\end{equation}
 taken with periodical boundary conditions. This definition brings the photon part of the Hamiltonian to a diagonal form. Here the bosonic commutators are 
\[
\left[b_{\vec{q},\lambda},b_{\vec{q}\;',\lambda'}^{+}\right]=\delta_{\vec{q},\vec{q}\;'}\delta_{\lambda\lambda'}
\]
and the unit vectors $\vec{e}_{\vec{q}}^{(\lambda)}$ are orthogonal  to the wave vector  $\vec {q}$ and to each other  
\[
\vec{q}\vec{e}_{\vec{q}}^{(\lambda)}=0\quad;\qquad \vec{e}_{\vec{q}}^{(\lambda)}\vec{e}_{\vec{q}}^{(\lambda')}=\delta_{\lambda\lambda'}\quad ;
\qquad\vec{e}_{\vec{q}}^{(\lambda)}=\vec{e}_{-\vec{q}}^{(\lambda)}\quad;\qquad(\lambda,\lambda'=1,2) \enspace .
\] 

With these ingredients and  normal ordering the operators  one gets the non-relativistic QED  Hamiltonian

\begin{eqnarray}
& & H^{QED}  =  \sum_{\vec{q},\lambda}\hbar\omega_{q}b_{\vec{q},\lambda}^{+}b_{\vec{q},\lambda}  \label{HQED}
\\
& &+\int\! d{\vec x} N\!\left[\frac{1}{2m}\!\left(\imath \hbar\nabla 
\psi^+({\vec x})
 +\frac{e}{c}({\vec A}_\bot({\vec x})+{\vec A}_{ext}({\vec x},t))\psi^+({\vec x})\right)\times \right.
\nonumber 
 \\
&&\left. \left(\frac{\hbar}{\imath}\nabla\psi({\vec x}, t) +\frac{e}{c}({\vec A}_\bot({\vec x})+{\vec A}_{ext}({\vec x},t)))\psi({\vec x})\right)\right]   \nonumber 
\\
&&+ \frac{1}{2}\int d\vec{x}\int 
 d{\vec x}'\psi^{+}(\vec{x})\psi^{+}(\vec{x}')\frac{e^{2}}{|\vec{x}-\vec{x}'|}\psi(\vec{x}')\psi(\vec{x}) + e\int d\vec{x} V_{ext}(\vec{x},t)\psi^+(\vec{x})\psi(\vec{x})\nonumber 
 \enspace .     \nonumber  
\end{eqnarray}

Here according to the general recipe of second quantization  a normal ordering  $N(... )$  had to be introduced also with respect to the photon creation and annihilation operators $b_{\vec{q},\lambda}^{+} $,  $b_{\vec{q},\lambda}$ and the photon frequency is $\omega_q= c|q|$. 
 
This non-relativistic QED Hamiltonian  coincides with the standard  one obtained directly  from the second quantized Hamilton operator of electrons interacting with a classical electromagnetic field in the Coulomb gauge, after the quantization of the transverse vector potential  according to  Eq. \ref{Aquant} and adding the energy of the photons, as it is given for example in \cite{Holstein}, while in most papers external potentials are not included. 

\begin{figure}[h]
\begin{center}
\includegraphics[scale=0.15]{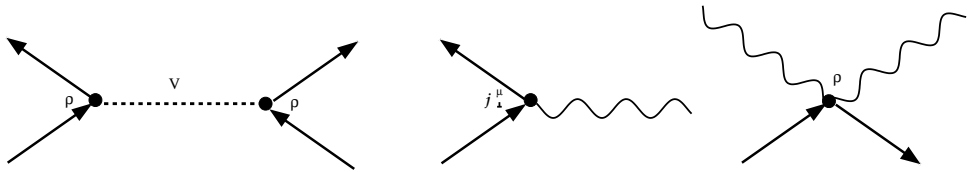}
\end{center}
\caption{The Feynman vertexes of the non-relativistic QED. The dashed line is the Coulomb potential, while the wavy line is the transverse photon.}
\label{vertexes}
\end{figure}

The interaction terms in the Hamiltonian  Eq. \ref{HQED} (in the absence of external fields!) are besides the usual Coulomb one of the many-body theories, a current-(transverse) field interaction  $\frac{1}{c} \vec{j}{\vec A}_\bot $  and a    "sea-gull" term  $\frac{e^2}{2m c^2} \psi^+ \psi {{\vec A}_\bot}^2$. The corresponding vertexes of the Feynman diagram technique within the adiabatic perturbation theory are illustrated on Fig. \ref{vertexes}. 

Obviously, this quantization procedure reintroduces electromagnetic  self-interaction on a hidden way. Its manifestation is the much discussed self-energy in the Feynman diagramms of the adiabatic perturbation theory.  This occurs also by the second-quantized version of the Coulomb interacting charged particles.  The Hamiltonian  in configuration space is not identical with the second quantized Coulomb Hamiltonian. Obviously their ground states (vacuum) are different.

The  Heisenberg equations of motion lead to the quantized (operator) coupled  Maxwell and Schr\"odinger equations. However their gauge invariant form does not imply  gauge invariance against arbitrary gauge transformation of the vector potential operator $\vec{A}(\vx)$. Gauge transformations like Eq.\ref{gauge} are meaningful only with  c-number functions 
$\chi(\vx, t)$  for the external field. 

We have shown, how a field theoretical treatment allows the use of the Lagrange formalism in deriving the non-relativistic quantum mechanical many-body theory of charged particles interacting with photons  avoiding the problems linked to point-like classical charges. Of course these results may be extended without problems to include also other charged particles, like the positive ions or holes  of the solid state theory.

\section{Ingredients from the relativistic  quantum field theory.}

Until now we described the QED of non-relativistic particles characterized by their mass and charge. However, from the relativistic quantum-field theory one knows, that these are not the only basic properties of elementary or composite particles. Although the non-relativistic QED is not reducible to the relativistic QED, one must take into account some ingredients one has learned from the quantum-field theory. At least the spin and the associated magnetic moment of electrons and nuclei (or ions) are indispensable  in a reasonable condensed matter theory. An essential role plays in this context the spin-statistic theorem of the relativistic theory stating that particles with integer spin are bosons, while those with half integer spin are fermions.

For this sake one has to include in the classical Maxwell equations Eq.\ref{Max1a} also a magnetic (or spin ) current

 \begin{equation}
\vec{j}^M(\vx)= c\nabla \times \vec{M}(\vx) \enspace .  
\end{equation}
The magnetization $\vec{M}(\vx)$ is given by the spin density. 
In the here discussed example of electrons with spin $1/2$ 
\begin{equation}
\vec{M}(\vx)=\frac{e\hbar}{2mc}\vec{\sigma}(\vx) \enspace ,
\end{equation}
with the spin density
\begin{equation}
 \vec{\sigma}(\vx)=\sum_{\sigma,\sigma'=\pm 1}\psi_\sigma(\vx)^*\hat{\vec{\sigma}}_{\sigma,\sigma'}\psi_{\sigma'}(\vx)
\end{equation}
and  the well-known $2\times 2$ sigma matrices $\hat{\vec{\sigma}}$.

To get the so extended Maxwell equations one has to add the piece
\begin{equation}
-\vec{B}(\vx)\vec{M}(\vx)
\end{equation}
to the Lagrangian density  Eq.\ref{Lagran} or Eq.\ref{Lagran1}, respectively with the opposite sign to the  Hamiltonians Eq.\ref{Hamiltoniana-Coul} and Eq.\ref{HQED}. The wave functions $\psi_\sigma (\vx)$ will then satisfy the Pauli equation
\begin{eqnarray}
\imath\hbar\frac{\partial}{\partial t}\psi({\vec x},t)&=&  \label{Pauli}
\left(\frac{1}{2m}\left(-\imath\hbar\nabla+\frac{e}{c}(\vec{A}(x,t)+\vec{A}_{ext})\right)^{2}\right. 
\\
&+&\left. e\left(V({\vec x},t)+ V_{ext}(\vec{x},t)\right)+\frac{e\hbar}{2 mc}\vec{B}(\vx)\hat{\vec{\sigma}}\right)\psi({\vec x},t) \enspace , \nonumber
\end{eqnarray}
with the  matrix-column notation 
\begin{equation}
\psi(\vx,t)=\left(\begin{array}{cl}\psi_+ (\vx)\\ \psi_-(\vx)\end{array}\right) \enspace .
\end{equation}

The added piece to the Lagrangians does not affect gauge-invariance and the quantization of this extended electromagnetic theory is trivial. Of course one has to take into account the connection between the spin and statistics (Fermi/Bose). 
Obviously,  the scheme we described in these chapters may be easily extended to any system of particles with arbitrary masses, charges and spins.

\section{Non-relativistic quantum-mechanical  many-body  Hamiltonian without photons including   $1\over c^2$ terms.}

Only quantum optics uses up to these days the above described non-relativistic QED of ensembles of charged particles and photons.  Many-body theories of solid-state or plasma restrict their task to the subspace of states containing no photons. In the description of electrical phenomena with longitudinal fields this theory hat enormous successes in understanding the properties of solids. However, the description of magnetic properties is not so consolidated, since the standard solid-state theory is still based just on Coulomb forces if we leave aside spin magnetism. 

The purpose of this Section is to derive from the non-relativistic QED the proper formulation of the $1/c^2$ many-body theory containing already the magnetic interactions between electrons. More than that is not possible without inclusion of the photons. 

To simplify the discussion we omit here the external fields and reintroduce them again at the end according to the "minimal rule" and denote by
\begin{equation}
\vec{i}(\vx,t)=\frac{e}{2m}\psi(\vx,t)^*\frac{1}{\imath} \hbar\nabla\psi(\vx,t) +c.c  
\end{equation}
the current density in the absence of a vector potential. 

We shall proceed in two steps: i) to look in the subspace of states without photons at the role of the different terms in the QED Hamiltonian we derived  and ii) to neglect retardation effects, that are not tractable with a local Hamiltonian in this restricted subspace and anyway give rise to terms of higher order as $1/c^2$. 

First we discuss the theory in  the absence of  external fields.
The terms that do not contain the transverse vector potential ${\vec A}_\bot$ remain unchanged. The term describing the energy of the free photons of course  has to be ignored. Then remains the Coulomb interaction and  three other sort of interaction terms we have to discuss. First at all the quadratic, so called "sea-gull" term  $\frac{e^2}{2m c^2} \psi^+ \psi {{\vec A}_\bot}^2$. It  may be also ignored, since it communicates with the photon vacuum only with the help of the terms $\frac{1}{c} \vec{i}{\vec A}_\bot $ and therefore may contribute only third order terms in $1/c$. Thus we remain only with these last interaction terms to discuss. 

\begin{figure}[h]
\begin{center}
\includegraphics[scale=0.3]{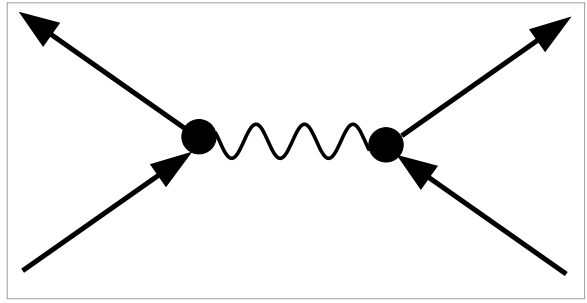}
\end{center}
\caption{The transverse photon exchange diagram in the QED. The wavy line represents the propagator of the transverse photon. }
\label{graph}
\end{figure}

We shall pursue the discussion within the frame of the adiabatic $S$-matrix theory (or of their Green functions). It is easy to see, that there  is only one basic graph and its combinations that may contribute by this vertex  to the $S$ matrix elements in the subspace of electron states. This is  shown in Fig. \ref{graph}. It describes the exchange of a transverse photon between two transverse electron currents mediated by the transverse photon propagator given 
 in the 4-dimensional Fourier space $\omega, {\vec q}$ by
\begin{equation}
\frac{1}{q^2-\omega^2 /c^2-\imath 0}(\delta_{\mu,\nu}-\frac{q_\mu q_\nu }{q^2})\quad ; \qquad (\mu,\nu=1,2,3)
\end{equation}
and the vertexes contain momentum factors corresponding to the  the currents (see Ref. \cite{Banyai1} )
 
After neglecting the term  $-\omega^2 /c^2 $  in the denominator (i.e. ignoring retardation) one  eliminates corrections of higher order as $1/c^2$ and  the  photon propagator looks as

\begin{equation}\frac{1}{q^2}(\delta_{\mu,\nu}-\frac{q_\mu q_\nu }{q^2})\quad ; \qquad (\mu,\nu=1,2,3)\enspace. 
\end{equation}
Since no pole survived, the $-\imath 0 $ term could have been also ignored and 
$\frac{4\pi}{q^2}$ is just the Fourier transform of the Coulomb potential.

Then one may convince oneself that these graphs coincide with the basic vertexes of the S matrix of an  $1/c^2$ purely electron Hamiltonian containing besides the kinetic energy and the Coulomb interactions the interaction term 
\begin{equation}
-\frac{1}{2} \int d\vec{x}\int d\vec{x'}
\frac{N\left[\vec{i}_\bot(\vec{x})\vec{i}_\bot(\vec{x}')\right]} 
{c^2|\vec{x}-\vec{x}'| }\enspace . \label{cur-cur}
\end{equation}
 This term has an appealing form analogous to the charge density-charge density Coulomb interaction.  It is the microscopical expression  of the important Biot-Savart law of  interaction between currents in finite macroscopic samples, where light propagation effects may be ignored. 
 
  One might argue that due to the smallness of the velocities in the condensed matter such an $1/c^2$ term may be neglected. This is obviously false. Our everyday experience teaches us, that a macroscopic number of slow electrons may create enormous magnetic fields.

 In order to keep  gauge invariance with respect to an external field $\vec {A}_{ext} $ one has to use again the minimal recipe and this requires also its introduction not only in the kinetic energy term, but also in the current-current interaction.  
  
This transverse current-current interaction Eq. \ref{cur-cur}  was derived in Ref. \cite{Lectures2},\cite{Banyai1}  within this QED frame. 

\begin{figure}[h]
\begin{center}
\includegraphics[scale=0.25]{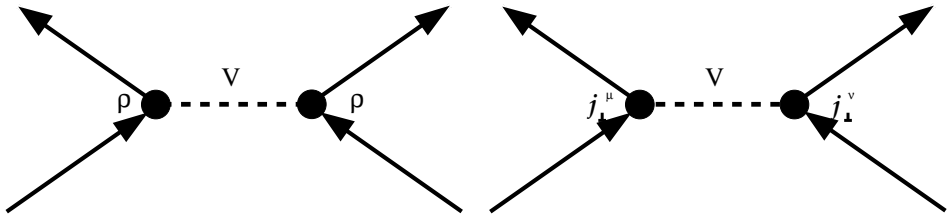}
\end{center}
\caption{The basic density-density and current-current vertexes  in the $1/c^2$ S-matrix. Here the dashed line indicates just a Coulomb potential. }
\label{diagram}
\end{figure}

Therefore, we may conclude that after reintroducing the external fields according to the minimal principle of Yang-Mills theories, the quantum mechanical $1/c^2$ electron Hamiltonian  that generates the diagrams of Fig.\ref{diagram} 

\noindent is just

 \begin{eqnarray}
{\bf H}(t)\!\! &=&\!\!\-\!\!\int\!\! d\vx \psi^{+}(\vx)\left[\frac{1}{2m}
\left(\frac{\hbar}{\imath}\nabla-\frac{e}{c}\vec{A}_{ext}(\vx,t)\right)^2 \!\!+\!\! eV_{ext}(\vx,t) \right]\psi(\vx) \label{Hc2}
\\
&+& \frac{1}{2} \int d\vx \int d\vxp
\frac{{\cal N}\left[\rho(\vx)\rho(\vxp)\right]}{|\vx-\vxp|} 
 -\frac{1}{2} \int d\vx\int d\vxp
\frac{{\cal N}\left[\vec{j}_\bot(\vx,t)\vec{j}_\bot(\vxp,t)\right]} 
{c^2|\vx-\vxp|}  \enspace .  \nonumber
\end{eqnarray}

Here $\rho(\vx)$ denotes the charge density operator
\begin{equation}
\rho(\vx)=e\psi^{+} (\vx)\psi(\vx) \enspace ,
\end{equation}
while $\vec{j}_\bot(\vec{x},t)$ denotes the transverse part of the current density operator in the presence of the external vector potential
\begin{equation}
\!\vec{j}(\vx,t)\!=\! \frac{e}{2m}\left(\!\psi^{+}(\vx)
\left(\frac{\hbar}{\imath}\nabla\!-\!\frac{e}{c}\vec{A}_{ext}(\vx,t)\!\!\right)\psi(\vx)\!+\!h.c.\right) \enspace .\label{curdens}
\end{equation}   

The continuity equation emerging from this charge conserving Hamiltonian is 
however
\begin{equation}
\nabla\vec{J}+\frac{\partial}{\partial t}\rho=0 \label{continuity1}
\end{equation}
with the conserved current
\begin{equation}
\vec{J}(\vx,t)=
\frac{e}{2m}\psi({\vec x},t)^*\left(-\imath \hbar\nabla
+\frac{e}{ c}({\vec A}_{ext}({\vx},t)+{\vec A}_\bot(\vx,t)) \right)\psi({\vx},t) +c.c \enspace ,
\end{equation}
where 
\begin{equation}
\vec{A}_{\bot}(\vx,t)\doteq \frac{1}{c}\int d\vxp \frac{\vec{j}_{\bot}(\vxp,t)}{|\vx -\vxp|}
\end{equation}
is (by neglecting the retardation!) the transverse vector potential generated by the electrons.

The generalization of these results  for a system of electrons and ions as constituents of the solid state is obvious.

%------------

This new  interaction in coordinate space, due to the additional integrals  in the definition of the transverse part and the normal products, is difficult to handle.  However, this term  has a simple expression in the discrete ${\vec k}$ - space basis (plane waves with periodical boundary conditions in a cube of volume $\Omega$). It looks explicitly as  
\begin{equation}
-\frac{e^{2}\hbar^{2}}{m^{2}c^2\Omega}\sum_{\sigma,\sigma'=\pm 1}\sum_{\vec{k},\vec{p},\vec{q}}\frac{2\pi}{q^{2}}\left(\vec{k}\vec{p}-\vec{q}\vec{k}\frac{1}{q^2}\vec{q}\vec{p}\right)a_{{\vec{k}},\sigma }^{+}a_{\vec{p},\sigma '}^{+}a_{\vec{p}+\vec{q},\sigma'} a_{\vec{k}-\vec{q},\sigma}\enspace .
\end{equation}
 %--------------------

For a better understanding of the underlying physics, let us consider now the Hartree approximation of this Hamiltonian. It looks as
 \begin{eqnarray}
{\bf H}_{Hartree}(t)\!\! &=&\!\!\-\!\!\int\!\! d{\vx} \psi^{+}(\vx)\left[\frac{1}{2m}
\left(\frac{\hbar}{\imath}\nabla-\frac{e}{c}\vec{A}_{ext}(\vx,t)\right)^2 \!\!+\!\!e V_{ext}(\vx,t) \right]\psi(\vx) \label{Hc2-Hart}
\\
&+&  \int d\vx \int d\vxp
\frac{\rho(\vx)\langle\rho(\vxp,t)\rangle}{|\vx-\vxp|} 
 -\int d\vx\int d\vxp
\frac{\vec{j}_\bot(\vx,t)\langle\vec{j}_\bot(\vxp,t)\rangle} 
{c^2|\vx-\vxp|}  \enspace , \nonumber 
\end{eqnarray}
where $\langle \rho(\vx,t)\rangle$ and $\langle {\vec j}_\bot(\vx,t)\rangle$
are the (chosen) ensemble averages of the charge and  transverse current densities.

One  may here identify the s.c. internal scalar and vector potentials 
\begin{equation}
V_{int}(\vx,t) =\int d\vxp\frac{\langle \rho(\vxp,t)\rangle}{|\vx -\vxp|}; \qquad
\vec{A}_{int}(\vx,t)=\int d\vxp\frac{\langle j_\bot(\vxp,t)\rangle}{c |\vx -\vxp|}\enspace . \label{mean}
\end{equation}
Of course, here in the definition of the internal vector potential the retardation is again missing! 

Eq.\ref{Hc2} and Eq.\ref{Hc2-Hart} together with the identifications of Eq.\ref{mean} show, that  the average Coulomb field created by the electron results from the  charge-charge interaction and  the average (dia-) magnetic field created by the velocity of the electron   results from the current-current interactions. This is analogous to the (ferro-) magnetic field of localized spins resulting from their mutual interaction in the Heisenberg spin model. 

One may rewrite Eq. \ref{Hc2-Hart} as
 \begin{eqnarray}
{\bf H}_{Hartree}(t) & = &-\int d{\vx} \psi^{+}(\vx)\left[\frac{1}{2m}
\left(\frac{\hbar}{\imath}\nabla-\frac{e}{c}\vec{A}_{ext}(\vx,t)\right)^2 +e V_{tot}(\vx,t) \right]\psi(\vx) \nonumber
\\
&-& \frac{1}{c}\! \int  d\vx \int 
\vec{j}_\bot(\vx,t)\vec{A}_{int}(\vx,t)  \label{Hc2-Hart1}
 \enspace . 
\end{eqnarray}

The above Hartree Hamiltonian coincides only up to second  order terms in the fields with the  mean field Hamiltonian resulted from the QED Hamiltonian Eq.\ref{HQED} by approximating  the total vector potential  ${\vec A}_\bot({\vec x})+{\vec A}_{ext}$ by his average $\vec{A}_{tot}$ and  the Coulomb term by its mean-field approximation.

\begin{equation}
H_{mean}=\int d{\vx} \psi^{+}(\vx)\left[\frac{1}{2m}
\left(\frac{\hbar}{\imath}\nabla-\frac{e}{c}\vec{A}_{tot}(\vx,t)\right)^2 \!\!+
\!\!e V_{tot}(\vx,t) \right]\psi(\vx)  \enspace .\label{meanvechi}
\end{equation}
 While the Hamiltonian Eq.\ref{Hc2-Hart1} is invariant only against time-independent gauge transformations of the external fields, the mean-field Hamiltonian Eq. \ref{meanvechi} is   formally gauge invariant against time independent  gauge transformations of the total field $\vec{A}_{tot}$ although actually the gauge of the internal vector potential ${\vec A}_{int}$ is already fixed to be the Coulomb one.

\section{Darwin's classical approach revisited. }
\label{sec:3}

In this  Section we show that Darwin's reasoning within the classical electromagnetic theory of point-like charges, if properly formulated in the Coulomb gauge i.e. considering only the true physical degrees of freedom of the magnetic field, leads to a classical  Hamiltonian analogous to the one we got in the previous Section.

As it is well-known, one can not formulate a Lagrangian theory of classical point-like charged particles interacting with the electromagnetic field due to the divergent self-interaction.( From the Lorentz force one has to omit the action of the field created by each charged particle on itself!) This impedes also a proper derivation of an $1/c^2$ Hamiltonian.

Almost one hundred years ago Darwin \cite{Darwin} proposed nevertheless a closed classical Lagrangian  for N point-like charges $e_i$ and mass $m_i$ ($i,j =1,\ldots N$) including terms up to order $1/c^2$ avoiding self-interaction and constructed the following Hamiltonian
\begin{eqnarray}
& &{\cal H}=\sum_{i}\frac{1}{2m_i}\vec{p_{i}}^{2}+\sum_{i>j}\frac{e_i e_j}{|\vec{r_{i}}-\vec{r}_{j}|} \label{Darwin}
\\
& &-\sum_{i>j}\frac{e_i e_j}{2c^{2}m_i m_j|\vec{r}_{i}-\vec{r}_{j}|}\left[\vec{p}_{i}\cdot\vec{p}_{j}+(\vec{p}_{i}\cdot\vec{n}_{ij})(\vec{p}_{j}\cdot\vec{n}_{ij})\right] ,
\nonumber
\end{eqnarray}
where $\vec{n}_{ij}\equiv \frac{\vec{r}_i -\vec{r}_j}{|\vec{r}_i -\vec{r}_j| }$. 
His derivation is based on the expansion of the Li\'enard-Wiechert potentials to second order in $\frac{1}{c}$.

The reasoning of Refs. \cite{Landau} and \cite{Jackson}  to derive the above Hamiltonian starts from the discussion of the potentials felt by one point-like particle in the field of another one at a distance $R$.
Landau and Lifshitz \cite{Landau}  use the Lorentz gauge, perform an expansion of the scalar potential up to second order in the finite distance between two particles divided by the light velocity ($R/c$)  and later  perform  a gauge transformation turning to the Coulomb gauge, without stating it explicitly.
 
 Jackson \cite{Jackson} starts  in the Coulomb gauge, gets the transverse current - transverse current Lagrange function for two particles  however performs an integration  by parts of the vector potential of the two particles at a finite distance leaving aside  the vanishing surface contribution and gets Darwin's Hamiltonian.
  
Thus the two approaches are similar and their results are identical. Moreover, from Jackson's derivation results that the transverse current - transverse current version is equivalent to Darwin's version whenever the distances between the particles are finite.

Without the mentioned partial integration Jackson's Hamiltonian \cite{Jackson} looks as 
\begin{eqnarray}
& & H =\sum_{i}\frac{\vec{p_{i}}^{2}}{2m_i}+\sum_{i>j}\frac{e_i e_j}{|\vec{r_{i}}-\vec{r}_{j}|}
\label{Hamclas}
\\
& &-\sum_{i>j}\frac{e_i e_j}{c^{2}m_i m_j}\vec{p}_i\left[
\frac{\vec{p}_j }{|\vec{r}_i-\vec{r}_j|}-\frac{1}{4\pi} \int d\vec{x}\frac{1}{|\vec{r}_i-\vec{x}|}\nabla \left(\vec{p}_j \nabla \frac{1}{|\vec{x}-\vec{r}_j|}\right)\right] \nonumber
\end{eqnarray}
and consists of transverse current - transverse current terms for $i\neq j$.

By introducing the charge and current densities:
\begin{equation}
\rho(\vec{x})=\sum_i e_i\delta(\vec{x}-\vec{r}_i); \qquad 
\vec{i}(\vec{x})=\sum_i \frac{e_i}{m_i}\vec{p}_i\delta(\vec{x}-\vec{r}_i)
\end{equation}
and ignoring the missing $i=j$ terms one might rewrite this classical Hamiltonia Eq.\ref{Hamclas}  as 

\begin{equation}
\sum_{i}\frac{1}{2m_i}\vec{p_{i}}^{2}+\frac{1}{2} \int d\vec{x}\int d\vec{x}'
\frac{\rho(\vec{x})\rho(\vec{x}')}{|\vec{x}-\vec{x}')|}  -\frac{1}{2} \int d\vec{x}\int d\vec{x'}
\frac{\vec{i}_\bot(\vec{x})\vec{i}_\bot(\vec{x}')} 
{c^2|\vec{x}-\vec{x}'|}\enspace ,\label{Hsymb}
\end{equation}
where 
$\vec{i}_\bot (\vec{x})$ is the transverse part of the current density
\begin{equation}
\vec{i}_{\bot}(\vec{r},t)\equiv\vec{i}(\vec{r},t)+\frac{1}{4\pi}\nabla\int d\vec{r}'\frac{\nabla'\vec{i}(\vec{r}',t)}{|\vec{r}-\vec{r}'|} \enspace .\label{c2-Hamil-cl}
\end{equation}

Under this form it looks similar to the result we obtained before (in the absence of other external fields).
A quantization of the classical Hamiltonian of the point-like charged particles Eq. \ref{Hamclas} is  then immediate by second quantization of Eq. \ref{Hsymb} as it was done in Refs. \cite{Lectures2}, \cite{Banyai1} avoiding the configuration space quantization  either of Eq. \ref{Hamclas} or Eq. \ref{Darwin}. This leads to the same Hamiltonian Eq. \ref{Hc2} as we deduced from the non-relativistic QED.

\section{The importance  of the diamagnetic  current-current interaction in many-body theories.}

\subsection{Superconductivity}
According to Bardeen, Cooper and Schrieffer \cite{BCS} the origin of the superconductive phase transition lies in the correlation between electrons of opposite momenta and spin resulted from  phonon exchange. 
However, to pursue this idea within the many-body theory of electrons interacting with phonons seems too difficult. Therefore, one tried to construct pure electron many-body theories with a built-in two-particle"attractive" potential $W(\vx)$  giving rise to no bound states, but just to  such correlations and implicitly to a superconducting phase transition. Such a model Hamiltonian is due to Rickayzen \cite{Rickayzen1,Rickayzen2} that we shortly describe here.  (The specific version of  Bogolyubov-de Gennes \cite{Bogo-deGenn} is included in the same frame.) The superconductive  phase transition is thought to be accompanied by a spontaneous  breaking of the phase invariance due to the non-vanishing anomalous average
$\langle\psi_{\frac{1}{2}}^{+}(\vec{x})\psi_{-\frac{1}{2}}^{+}(\vec{x'})\rangle$ .

Rickayzen's self-consistent Hamiltonian for a homogeneous electron system in the presence of a magnetic field described by the potential vector  ${\vec{\cal A}}({\vec x})$   is 
\begin{eqnarray}
& &{\cal H}_{s.c.}  
= \sum_{\sigma=\pm\frac{1}{2}}\!\!\!\int d\vx \psi_{\sigma}^{+}(\vx)\!\bigg\{ \frac{1}{2m}\left(-\imath\hbar\nabla-\frac{e}{c}\vec{\cal{A}}(\vx)\right)^{2} -\mu  \bigg\} \psi_{\sigma}(\vx) \label{HFB}  
\\
 & &+\frac{1}{2}\int d\vx\int d\vxp W(\vx-\vxp)\left[ \langle\psi_{\frac{1}{2}}^{+}(\vx)\psi_{-\frac{1}{2}}^{+}(\vxp)\rangle\psi_{-\frac{1}{2}}(\vxp)\psi_{\frac{1}{2}}(\vx)\right.
\nonumber
 \\
 &&+ \left.\langle\psi_{-\frac{1}{2}}(\vxp)\psi_{\frac{1}{2}}(\vx)\rangle\psi_{\frac{1}{2}}^{+}(\vx)\psi_{-\frac{1}{2}}^{+}(\vxp)
-\langle\psi_{\frac{1}{2}}^{+}(\vx)\psi_{-\frac{1}{2}}^{+}(\vxp)\rangle\langle\psi_{-\frac{1}{2}}(\vxp)\psi_{\frac{1}{2}}(\vx)\rangle\right]  \enspace . \nonumber
\end{eqnarray}

One can show that in the absence of the field ${\vec{\cal A}}({\vec x})$ a phase transition may occur below a critical temperature, provided the potential $W({\vec x})$ ensures a non-vanishing solution for the symmetry breaking gap parameter of the largely described  "gap equation" we do not give here. This condition is equivalent to the vanishing of the first derivative of the free energy with respect to the same parameter.

In order to prove the ideal diamagnetism (Meissner effect) one uses equilibrium  linear response. Here it is essential to understand the magnetic perturbation.
In the Hamiltonian of Eq.\ref{HFB} the (transverse) vector potential $\vec{\cal{A}}$ is a classical field that was introduced just by the  minimal principle, without any deeper justification. 

According to our previous Sections (see Eqs. \ref{Hc2-Hart} and \ref{mean} ) the correct s.c. Hamiltonian \cite{banyai-epjb} derived from the QED in the absence of photons is quite different, namely the kinetic energy term of Eq. \ref{HFB} in the presence of a transverse external vector potential is replaced by 
\begin{equation}
 \psi_{\sigma}^{+}(\vx)\! \frac{1}{2m}\left(-\imath\hbar\nabla-\frac{e}{c}\vec{A}_{ext}(\vx)\right)^{2}\psi_\sigma (\vx) -
\frac{1}{c}\vec{A}_{int}(\vx)\vec{j}_\bot (\vx) \enspace .
 \end{equation}
 
 In the linear approximation with respect to $\vec{A}_{ext}$ one gets the magnetic perturbing term 
\begin{equation}
-\frac{1}{c}\big(\vec{A}_{ext}(\vx)+\vec{A}_{int}(\vx)\big)\vec{j}_\bot(\vx)=
-\frac{1}{c}\vec{A}_{tot}(\vx)\vec{j}_\bot (\vx)
 \enspace .
\end{equation} 

Therefore, at least at the level of the linear response one should identify  the vector potential  $\vec{\cal{A}}$ not with the external but with total one. 

Taking also into account the peculiarity of the linear response within self-consistent theories that the deviation of the averages from their equilibrium values constitute so called  induced perturbations one may derive the relation between the average induced current and this total field (in  Fourier transforms) of the two transverse vectors reads as
\begin{equation}
\langle\tilde{j}_{\mu}(\vk)\rangle=\kappa(k)\tilde{{\cal A}}_{\mu}(\vk) \quad ;
\qquad (\mu=1,2,3)
\label{response}
\end{equation}
with the scalar  coefficient $\kappa(k)$ in an infinite homogeneous, isotropic  system  being a function of  $k=\sqrt{{\vk}^2 }$. Although this coefficient cannot be calculated explicitly,  it has been proven \cite{Banyai-Gartner} that for any potential $W(\vx)$ that leads to a stable superconductive phase in the absence of a magnetic field the coefficient $\kappa(0)$ is finite. This was early stipulated by Schafroth \cite{Schafroth} as being the necessary condition for superconductivity.

Thus we may conclude that
\begin{equation}
\langle \vec{\tilde{j}}(\vk)\rangle=\kappa(k)\big(
\vec{\tilde{ A}}_{ext}(\vk)+ \vec{\tilde{ A}}_{int}(\vk)\big)\quad ;
\qquad (\mu=1,2,3) \enspace .
\label{response1}
\end{equation}

Now, in the absence of retardation we have the relation of the internal vector potential to the average transverse current density:
\begin{equation}
\vec{\tilde{A}}_{int}(\vk)=\frac{4\pi }{ck^2 }\langle \vec{\tilde{j}}_{\bot}(\vk)\rangle \enspace . \label{internfield}
\end{equation} 
Using this relationship in the previous equation we get

\begin{equation}
\tilde{A}_{int}^{\mu}(\vk)=\frac{\frac{4\pi}{ck^2}\kappa(k)}{1-\frac{4\pi }{ck^2 }\kappa(k)}\tilde{A}_{ext}^{\mu}(\vk)\quad ; \qquad (\mu=1,2,3)\enspace .
\end{equation}

For the internal and external magnetic fields holds  similarly 

\begin{equation}
{\tilde B}^{\mu}_{int}(\vk)=\frac{\frac{4\pi }{ck^2 }\kappa(k)}{1-\frac{4\pi }{ck^2 } 
\kappa(k)}{\tilde B}^{\mu}_{ext}(\vk)\quad; \qquad (\mu=1,2,3)\enspace .\label{B-comp}
\end{equation}
This last equation has been obtained also by Tinkham  {\cite{Tinkham} assuming the interpretation of the vector potential 
 $\vec{\cal{A}}$ in Eq. \ref{Hc2-Hart} as being the total vector potential as in Eq. \ref{meanvechi}.
   Although it was just an educated  guess, it occurred to be correct within the linear approximation. The same is true about the Ginzburg-Landau non-linear theory of superconductivity \cite{Ginz-Land}.
   
   The same relation Eq. \ref{B-comp} may be obtained through Zubarev's \cite{Zubarev} early reasoning
 based just on the macroscopic Maxwell equations, without any reference to a Hamiltonian. 
  
Since $\kappa(0)\neq 0$, from Eq. \ref{B-comp} follows 
\begin{equation}
{\tilde B}^{\mu}_{int}(0)=-{\tilde B}^{\mu}_{ext}(0)\enspace 
\end{equation}
 i.e. a homogeneous external magnetic field is perfectly compensated by the internal magnetic field of the electrons. This shows that at least against a homogeneous magnetic field the system behaves as an ideal diamagnet.
 
As we have seen, although the correct electromagnetic formulation at the level of the linear response did not predict unexpected results, it eliminated some conceptual confusions. On the other hand, by strong magnetic fields, where linear response does not work, the correct $1/c^2$ Hamiltonian Eq.\ref{Hc2} may be very important.

\subsection{Linear response and thermal noise}

\subsubsection{Linear response of charged particles.}
Non-equilibrium linear response and the related problem of thermal fluctuations is another field  where the proper treatment of the magnetic field within many-body theory is important.  

Originally  Kubo \cite{Kubo} developed adiabatic linear response only with respect to a homogeneous electric field. However, it may be extended within the non-relativistic QED to a general external electromagnetic perturbation  
 due to  applied electromagnetic scalar $V^{ext}(\vx,t)$ and vector potentials
 $\vec{A}^{ext}(\vx,t)$ 
 
\begin{equation}
H'(t)=\int d\vx\left\{ \rho(\vx)V^{ext}(\vx,t)-
\vec{j}(\vx)\vec{A}^{ext}(\vx,t)\right\} \enspace ,\label{perturb}
\end{equation}
the average of the current density operator $\vec{j}(\vx,t)$ being given by 
\begin{equation}
\langle j_{\mu}(\vx,t)\rangle \!=\!\!\!\lim_{s\to+0}\int_{-\infty}^{t} \!\!\!\!\!dt'e^{s t'}
\!\int_{0}^{\beta}
\!\!\!\!d\lambda \!\int\!\!\! d\vxp \langle j_{\nu}(\vxp,-\imath\hbar\lambda)
j_{\mu}(\vx,t-t')\rangle_0 E_{\nu}^{ext}(\vxp,t') \enspace ,\label{kubo-ext}
\end{equation}
where  $\mu,\nu=1,2,3$ are vector-indices and summation over double indices is understood, while $s$ is the adiabatic parameter we shall omit for simplicity  in the following. 

In translation and rotation (in space) invariant systems one has in Fourier transforms a local relationship 
\begin{equation}
\langle{\tilde j}_{\mu}(\vk,\omega)\rangle= \kappa(\vk,\omega)_{\mu\nu} \tilde{E}^{ext}_\nu (\omega,\vk) 
\end{equation}
 with 
\begin{equation}
 \kappa(\vk ,\omega)_{\mu\nu}=\int_{0}^{\infty}
\!\!\!dt\int_{0}^{\beta}\!\!\!d\lambda\int \!\!\! d\vx e^{\imath(\vk\vx+\omega t)}
\langle j_{\nu}(0,0) j_\mu(\vx,t+\imath\hbar\lambda )\rangle_0 \enspace .
\label{susc}
\end{equation}

If the system under consideration is isotropic, then one may separate the longitudinal and transverse parts

\begin{equation}
 \kappa(\vk, \omega)_{\mu\nu}=\frac{k_{\mu}k_\nu}{k^2}\kappa_L(k,\omega) +
\left(\delta_{\mu\nu}- \frac{k_{\mu}k_\nu}{k^2}\right)\kappa_T(k,\omega) \enspace .
 \end{equation}

It is important to remark here that the frequency and the wave vector in the transverse case are not independent ($\omega=ck$). 

The next important point, first remarked by Izuyama \cite{Izuyama} and \cite{Zubarev} is  that $\kappa(\vk)$ does not coincide with the complex conductivity $\sigma(\omega, \vk)$, since this one characterizes the response to the total electric field in matter to which the electrons also contribute.
This total electric field is given by adding the average internal field to the external field
\begin{equation}
\vec{E}(\vk,t)=\vec{E}^{ext}(\vk,t)+\vec{E}^{int}(\vk,t) \enspace .
\end{equation}

According to the Maxwell equations of the non-relativistic QED the Fourier transforms of the internal longitudinal and transverse electric fields  are given by

\begin{equation}
\imath \vk\vec{\tilde{E}}_L (\vk,t)=\frac{4\pi}{k} \tilde{\rho}(\vk,t)\label{EL}
\end{equation}
and
\begin{equation}
\vec{E}_T(\vx,t)=-\frac{1}{c}\frac{\partial}{\partial t}\vec{A}_T (\vx,t) \enspace ,
\end{equation}
with the transverse vector potential operator (radiation field) given by  Eq. \ref{Aquant}. After neglecting retardation 
the transverse electric field may be expressed directly through the transverse current density 
\begin{equation}
\vec{ \tilde{E}}_T(\vk,t)=  -\frac{4\pi}{(ck)^2}\frac{\partial}{\partial t} \vec{\tilde{j}}_T (\vk,t) \label{ET} \enspace .
\end{equation}
The internal fields are then defined by the average charge , respectively current densities

\begin{eqnarray}
\imath \vk\vec{\tilde{E}}_L^{int} (\vk,t)&=&\frac{4\pi}{k} \langle \tilde{\rho}(\vk,t)\rangle
\label{intL}
\\
\vec{\tilde{E}}_T^{int}(\vk,t)&=&  -\frac{4\pi}{(ck)^2}\frac{\partial}{\partial t} \langle \vec{\tilde{j}}_T (\vk,t)\rangle \enspace .\label{intT}
\end{eqnarray}

As a consequence the longitudinal and transverse conductivities are related to the coefficients $\kappa_{L/T}(k,\omega)$ (see \cite{Banyai-Bundaru-Gartner})} by
\begin{equation}
\sigma_L(k,\omega)=\frac{\kappa_L(k,\omega)}{1-\imath\frac{4\pi}{\omega}\kappa_L(k,\omega)} \enspace , \label{sigmaL}
\end{equation}
respectively
\begin{equation}
\sigma_T(k,\omega)=\frac{\kappa_T(k,\omega)}{1+\imath\frac{4\pi}{c^2k^2}\kappa_T(k,\omega)} \enspace . \label{sigmaT}
\end{equation}
Since the transverse external electric field obeys the wave equation, in the last equation the variables $\omega$ and $k$ are related by $\omega=ck$ ! Therefore one might speak only about a  frequency dependent transverse conductivity 
\begin{equation}
\sigma_T(\omega)=
\frac{\kappa_T(\omega/c, \omega)}{1+\imath  \frac{4\pi }{\omega}\kappa_T(\omega/c,\omega)} \enspace .\label{sigma-kappaom}
\end{equation}

In Eq. \ref{ET} relating  the internal transverse electric field to the average transverse electronic current we already neglected the retardation, therefore it is consequent to neglect it also in calculating the coefficient   
$\kappa_T(k,,\omega)$ within the $1/c^2$ Hamiltonian of Eq. \ref{Hc2} (including possible other terms due to  interaction with phonons or impurities).

\subsubsection{Electromagnetic thermal noise.}
There is an interesting relationship 
between these conductivities given by the consequent linear response theory of charged particles and their thermal noise. Such a connection was first shown in a fluctuation-dissipation theorem by Callen and Welton \cite{Callen-Welton}. However, in their derivation the electromagnetic  field of the charged particles was ignored. 
A proper electromagnetic analysis for Coulomb interacting particles was performed in Ref. \cite{Banyai-Aldea-Gartner} and recently \cite{Banyai-Bundaru-Gartner} extended to include the transverse case with diamagnetic interaction.

The time fluctuation $\Delta_X (t)$ of a given observable (hermitian operator) $X$ in thermal equilibrium is defined \cite{Lifshitz-Pitaevskii} as the average square deviation
\begin{equation}
\Delta_X(t)=\langle\left(X(t)-X(0)\right)^2\rangle_0 \geq 0 \enspace , \label{defin}
\end{equation}
where the average is taken over the macro-canonical equilibrium density matrix.
This is analogous to the general definition of a fluctuation both in classical- or quantum-statistics.

Leaving apart the constant $2\langle X(0)^2\rangle$ usually chosen to be vanishing, the entity of interest is
\begin{equation}
\delta_X(t)=\langle X(t)X(0)+X(0)X(t)\rangle_0  \enspace ,
\end{equation}
which is a real  and even function of $t$. As a consequence its Fourier transform
\begin{equation}
 {\tilde{\delta}_X}(\omega)=\int_{-\infty}^\infty dt e^{\imath \omega t}\delta_X(t)= 2\int_0^\infty dt \cos(\omega t)\delta_X(t) \enspace  \label{spectrum}
\end{equation}
is also real and even. Moreover, according to the Wiener-Khinchin Theorem \cite{Wiener, Khinchine}, \cite{deGroot} it is positive.
This is  defined \cite{Lifshitz-Pitaevskii} as the "noise" spectrum of $X$ in quantum statistics

It is easy to show, by expansion in the basis of the eigenfunctions of the Hamiltonian  for any observable $X$ (here $X\equiv X(0)$) the following identity
\begin{equation}
{\tilde{\delta}_X}(\omega)=
2\hbar \omega \coth(\frac{\beta\hbar \omega}{2})\Re\int_0^{\infty}dt
 e^{-\imath \omega t}
 \int_0^{\beta}d\lambda
 \langle XX(t+\imath \hbar \lambda\rangle_0 \enspace . \label{noise2}
\end{equation}

As we have seen, this kind of correlators appear in the linear response theory.

Now, let us consider the  noise  spectrum  of an electric field. In Fourier transforms it is easy to define  the scalars characterizing the longitudinal, respectively its transverse components 
$E_L (\vk)\equiv\frac{\vk}{k} \vec{\tilde{E}}(\vk,t)$ and
$E_{T(\lambda)}\equiv \vec{e_\lambda}\vec{\tilde{E}}(\vk,t)$ with 
$\vec{e}^{(\lambda)}_\vk=
\vec{e}^{(\lambda)}_{-\vk}$ ($\lambda=1,2 $)
 being the two unit vectors defining the independent polarizations of the transverse field.

From the hermiticity of the operator $\vec{E}(\vx,t)$ it follows that its Fourier transform obeys $\vec{\tilde{E}}(-\vk,t)^+ =\vec{\tilde{E}}(\vk,t)$. We  may define  however 
 two hermitian scalar operators (observables) as the "real" and "imaginary" parts of the operators. In what follows we shall omit the index $\lambda$ since due to isotropy neither the proof nor the results depend and consider the noise of the observables 
 \begin{eqnarray}
E_L^R(\vk,t)&=&\frac{\vk}{2k} (\vec{{\tilde E}}(\vk,t)+
\vec{{\tilde E}}(-\vk,t))
\\
E_L^I(\vk,t)&=&\frac{\vk}{2k\imath} (\vec{{\tilde E}}(\vk,t)-
\vec{{\tilde E}}(-\vk,t)) \enspace ,
\end{eqnarray}
 respectively
 \begin{eqnarray}
E_T^R(\vk,t) &=&\frac{1}{2} \vec{e}_\vk( \vec{\tilde{E}}_T(\vk,t)+ \vec{\tilde{E}}_T(-\vk,t))
\\
E_T^I(\vk,t) &=&\imath\frac{1}{2\imath} \vec{e}_\vk( \vec{\tilde{E}}_T(\vk,t)-
\vec{\tilde{E}}_T(-\vk,t)) \enspace .
\end{eqnarray}

Then we use  Eqs. \ref{EL}, \ref{ET} that express these operators through the charge density and current density operators while neglecting the retardation and
consequently  considering the time evolution  also in the $1/c^2$ approximation i.e. 
in the frame of the Hamiltonian  Eq. \ref{Hc2}. 

Starting from definition of the above defined hermitian field operators and    of the noise spectral density Eq. \ref{spectrum}, after some algebra using translation and rotation  invariance in the coordinate space, as well as some partial integrations one  gets for the longitudinal and transverse noise spectra relationships to the complex conductivities defined by Eqs. \ref{susc},  \ref{sigmaL}, \ref{sigma-kappaom}
\begin{eqnarray}
{\tilde \delta}_{E_L}(\omega,k)
&=&-8\pi\Omega\hbar \coth(\frac{\beta\hbar \omega}{2})\Im{\frac{1}{1+\frac{4\pi\imath}{\omega}\sigma_L(k,\omega)}} \nonumber
\\
&=&-8\pi\Omega\hbar \coth(\frac{\beta\hbar \omega}{2})\Im{\frac{1}{\epsilon_L(k,\omega)}} \enspace ,
 \label{Lnoise}
\end{eqnarray}
respectively
\begin{equation}
{\tilde \delta}_{E_T}(\omega,k)|_{k=\omega/c}= 8\pi \Omega\hbar
\coth(\frac{\beta\hbar \omega}{2})\Im \frac{1}{1-\frac{4\pi\imath}{\omega}\sigma_T(\omega)}
\label{Tnoise}
\end{equation}
In Eq.\ref{Lnoise} we used the relation between the complex dielectric function and the complex conductivity
\begin{equation}
\epsilon(k,\omega)=1+\frac{4\pi\imath}{\omega}\sigma(k,\omega)
\end{equation}
and the inverted order of arguments $(\omega,k)$ in the notation underlines that $\vk$ is the wave-vector of the electric field, while $\omega$ is the frequency of the noise spectrum.  

These are the most general results for the noise spectrum of the electric field of wave vector $\vk$.
From  Eqs. \ref{Lnoise},  \ref{Tnoise} and the positivity of the noise spectral density follows also the positivity of the real part of the longitudinal/transversal conductivity ($\Re \sigma(\omega,k)\geq 0$) within the frame of the linear response theory.

 In the $\hbar\to 0$ limit one gets the results of the classical plasma theory \cite{Burgess}.

In the peculiar case of a homogeneous in space  longitudinal electric field ${\cal E}(t)$ (having a Fourier transform in space proportional to the volume $\Omega=LS$) one gets for the potential drop $U(t)={\cal E}(t)L$ the noise spectral density

\begin{equation}
{\tilde \delta}_{U}(\omega,0)
=-8\pi\hbar\frac{L}{S} \coth(\frac{\beta\hbar \omega}{2})\Im{\frac{1}{1+\frac{ 4\pi \imath}{\omega}\sigma_L(\omega)}} \label{eq:Unoise}
\end{equation}

Now, one has the frequency dependent resistance 
\begin{equation}
R(\omega)=\frac{L}{S\Re \sigma_L(\omega)} \label{resistance}
\end{equation}
along the $z$ direction  and one may define a capacity between the end cross-sections of the sample  

\begin{equation}
C(\omega)=\frac{S}{L 4\pi}\Re \epsilon_L(\omega)=\frac{S}{L 4\pi}(1-\frac{4\pi}{\omega}\Im \sigma_L(\omega)) \enspace. \label{capacity}
\end{equation}
If the resistance and the capacitor are parallel linked, then the resulting impedance $Z(\omega)$ is
\begin{equation}
\frac{1}{Z(\omega)}=\frac{1}{R(\omega)}+\imath\omega C(\omega) \label{impedance}
\end{equation} 
then one gets from Eq.\ref{eq:Unoise}
\begin{equation}
{\tilde \delta}_{U}(\omega,0) =2\hbar \omega \coth(\frac{\beta\hbar \omega}{2})\Re Z(\omega)=
2\hbar \omega \coth(\frac{\beta\hbar \omega}{2})\frac{R(\omega)}{1+(\omega R(\omega)C(\omega))^2}
\enspace .
\end{equation}
 
For vanishing noise frequency  
\begin{equation}
{\tilde \delta}_{U}(0,0)=4R k_{B}T \label{Nyquist0}
\end{equation}
and one recovers the old Nyquist theorem \cite{Nyquist} .

Beside the field noises one might consider also the noise of the photon occupation numbers.  The noise spectral density of the photon occupation  numbers $n_{\vk}$ (with an  arbitrary chosen polarization not mentioned) is given by
  
\begin{equation}
\tilde{\delta}_{n_{\vk}}(\omega)
=2\hbar \omega \coth(\frac{\beta\hbar \omega}{2})\Re\int_0^{\infty}dt
 e^{-\imath \omega t}
 \int_0^{\beta}d\lambda
 \langle {\hat n}_{\vk}{\hat n}_{\vk}(t+\imath \hbar \lambda\rangle_0 \enspace , \label{pnoise}
\end{equation}
where ${\hat n}_{\vk}\equiv b_{\vk}^+ b_{\vk}$ is the operator of the number of photons of wave vector $\vk$ and the given polarization  we do not mention in the notation.  

 On the other hand, by two partial integrations (using time translation invariance and the assumed decay of the correlations) one may show that
\begin{equation}
\int_0^{\infty}\!\!\!\!\!dt
 e^{-\imath \omega t}\langle \hat{n}_{\vk}(0) \hat{n}_{\vk}( t+\imath \hbar \lambda)\rangle_0=
\frac{1}{\omega^2} 
 \int_0^{\infty}\!\!\!\!\!dt
 e^{-\imath \omega t}\langle \dot{\hat {n}}_{\vk}(0)\dot{\hat{n}}_{\vk}( t+\imath \hbar \lambda)\rangle_0  \enspace .
\end{equation} 
 
 However, retaining the lowest order in $1\over c$ from the full QED Hamiltonian
 \begin{equation}
\frac{1}{c}\vec{j}_T(\vx,t)\vec{A}_T
\end{equation} 
we have 
\begin{equation}
{\dot n}_{\vk}(t)=\frac{\imath}{\hbar c}\int\!\! d\vx \big[\vec{j}_T(\vx,t)\vec{A}_T(\vx,t),n_{\vk}(t)\big]
\end{equation}
and
\begin{eqnarray}
&&\int_0^{\infty}\!\!\!\!\!dt
e^{-\imath \omega t}\langle n_{\vk}(0) n_{\vk}( t+\imath \hbar \lambda)\rangle_0 =
-\frac{1}{(\hbar\omega c)^2} 
 \int_0^{\infty}\!\!\!\!\!dt
 e^{-\imath \omega t}
 \int\!\! d\vx \int\!\! d\vxp
 \\
&& \langle\big[ \vec{j}_T(\vx,0)\vec{A}_T(\vx,0),n_{\vk}(0)\big]
\big[\vec{j}_T(\vxp,t+\imath\hbar\lambda)\vec{A}_T(\vxp,t+\imath\hbar \lambda),n_{\vk}(t+\imath \hbar \lambda)\big]  
 \rangle_0  \enspace , \nonumber
\end{eqnarray} 
where the transverse  vector potential operator  is defined by Eq. \ref{Aquant}.

In what follows one may  neglect as before retardation and ignore consequently terms higher order than $1/c^2$. Within this approximation one remains only with an approximate Hamiltonian being the sum of the $1\over c^2 $ e.m Hamiltonian Eq. \ref{Hc2} (as well as some other interactions) for electrons and of the free Hamiltonian of the photons
both in the averaging over equilibrium as in the time evolution.

Performing the commutations and taking again into account the translation, rotation and reflection invariance (in the coordinate space) we get after some algebra 
\begin{eqnarray}
&&\tilde{\delta}_{n_{\vk}}(\omega)=2\frac{\coth(\frac{\beta\hbar\omega}{2})}{ck\omega}
\Re\!\!\!\int_{0}^{\infty}\!\!\!\!\!\!\!dt e^{-\imath \omega t}\!\!\!\int_{0}^{\beta}\!\!\!\!\!\!d\lambda\langle\tilde{j}(\vk,0)j_{T}(0,t+\imath\hbar\lambda)\rangle_{0}^{e}\times \nonumber
\\
&&\big[e^{-\imath(\omega-ck)t-\ ck\lambda}(1+{\cal N}_k)+e^{-\imath(\omega+ck)t+ ck\lambda)}{\cal N}_k\big] \enspace , \label{nnoise}
\end{eqnarray}
where 
\begin{equation}
{\cal N}_k\equiv\frac{1}{e^{\beta kc}-1} 
\end{equation}
is the Bose distribution of photons.

 In the  above expression again the electronic  current-current correlator appears  as in the case of the noise of the transverse electric field, however they differ essentially due to the presence of the Bose functions  under the integrals.  Therefore, it cannot be related to the transverse conductivity. Nevertheless, Eq. \ref{nnoise} may be the starting point for the direct computation of the photon number noise spectrum within given solid state models or mesoscopic systems. 
 
The thermal noise spectra of the longitudinal and transverse electric fields together with this last discussed case of photon number fluctuations exhaust the possible thermal noise measurements

\section{Conclusions.}
 We have discussed in some details the implementation of electromagnetism in quantum mechanics. The purpose of this analysis was to avoid the misinterpretations persistent in the treatment of the magnetic field in solid state theory. The basic ideas are length well-known, nevertheless there is a total confusion in the literature  about the distinction between the magnetic field in the matter and the applied classical macroscopic (external)  magnetic field. Another misunderstanding even at the textbook level is about gauge transformations in the Hamiltonian. We stress also the essential role of the classical, macroscopic external fields in the description of any electromagnetic experiment.

In order to bring clarity in this important matter one has to drop the old-fashioned  presentation of the quantum mechanics of electromagnetically interacting charged particles based on the classical theory of point-like charges. One must choose the formalism  of the quantum field theory in a non-relativistic frame. This is mandatory, since in condensed matter one has to do with ions (or nuclei) that are not elementary particles, therefore one cannot resort to the relativistic QED. The non-relativistic QED was length used in quantum optics, but not yet in many-body theories of condensed matter. An intermediary role might play the $1/c^2$ approximation of this non-relativistic QED that separates the motion of the electrons from that of the photons. The so obtained Coulomb gauge Hamiltonian includes the microscopic variant of the Biot-Savart law and is in concordance with the 100 year old proposal of Darwin \cite{Darwin}. 
We show, that a proper discussion of superconductivity and of the electromagnetic noise spectra may not ignore this ingredient.

%\vspace{1cm}
%{\Large {\bf Acknowledgement.}}

\end{document}